\documentclass[conference]{IEEEtran}

\usepackage{enumitem}
\usepackage{graphicx}
\usepackage{amsmath}
\usepackage{xcolor}
\definecolor{teal}{HTML}{008080}
\usepackage[
    colorlinks=true,
	citecolor=teal,
	linkcolor=teal,
	urlcolor=teal]{hyperref}
\usepackage{mathtools}
\usepackage{makecell}
\usepackage{booktabs}
\usepackage{multirow}
\usepackage{siunitx}

\newcommand{\bts}[1]{{\bf{#1}}}
\newcommand{\header}[1]{\textbf{#1}}
\newcommand{\device}[1]{{\texttt{ibm$\_$#1}}}
\newcommand{\actspace}[2]{{$(#1 e,#2 o)$}}
\newcommand{\scinot}[2]{#1 \cdot 10^{#2}}
\newcommand{\RIKEN}{RIKEN Center for Computational Science}
\newcommand{\CCF}{Cleveland Clinic Foundation}
\newcommand{\YKT}{IBM Quantum, T.J. Watson Research Center}
\newcommand{\JPN}{IBM Quantum, IBM Research -- Tokyo}

\newif\ifshowfig
\showfigtrue

\ifshowfig%
\else
  \renewcommand{\includegraphics}[2][]{}
\fi

\begin{document}

\title{Crossing the 12,000-atom barrier with heterogeneous quantum-classical supercomputing: quantum chemistry of protein-ligand complexes}

\author{
\IEEEauthorblockN{Kenneth M. Merz, Jr., Akhil Shajan, Danil Kaliakin, Fangchun Liang}
\IEEEauthorblockA{\CCF \\ \{merzk, shajana, kaliakd, liangf\}@ccf.org} \\
\IEEEauthorblockN{Yuichi Otsuka, Tomonori Shirakawa, Lukas Broers, Han Xu, Miwako Tsuji, Mitsuhisa Sato, Seiji Yunoki}
\IEEEauthorblockA{\RIKEN \\ \{otsukay, t-shirakawa, lukas.broers, han.xu, miwako.tsuji, msato, yunoki\}@riken.jp}  \\
\IEEEauthorblockN{Ryo Wakizaka, Yukio Kawashima, Jun Doi, {Hitomi Takahashi,} Toshinari Itoko, Hiroshi Horii}
\IEEEauthorblockA{\JPN \\ \{ryo.wakizaka, yukio.kawashima\}@ibm.com, \{doichan, {hitomi,} itoko, horii\}@jp.ibm.com} \\
\IEEEauthorblockN{\hspace{-0.7cm}Thaddeus Pellegrini, Javier Robledo Moreno, Kevin J. Sung, Ella Fejer, Robert Walkup, Seetharami Seelam, Mario Motta}
\IEEEauthorblockA{\YKT \\ \{Thaddeus.Pellegrini, j.robledomoreno, kevinsung, ella.fejer, mario.motta\}@ibm.com, \{walkup, sseelam\}@us.ibm.com} \\
}

\maketitle

\begin{abstract}
We develop a workflow decomposing a molecule into fragments via quantum embedding and simulating them with a heterogeneous quantum-classical (HQC) method. We sample fragment electronic configurations on two 156-qubit quantum processors (\device{cleveland}, \device{kobe}), using up to 94 qubits, running {21,006} circuits for over {239} hours, collecting {$\scinot{3.0}{9}$} measurement outcomes — the most resource-intensive HQC computation for quantum chemistry to date. We compute fragment wavefunctions via optimized subspace diagonalization on {supercomputers Fugaku, Miyabi-G, and ROQUO}, achieving 72.5$\%$ parallel efficiency with scalable distributed linear algebra kernels. We simulate two protein-ligand complexes spanning dispersion- and electrostatics-dominated regimes (11,608, 12,635 atoms), demonstrate $>40\times$ increase in system size and up to $210\times$ improvement in accuracy over previous state-of-the-art, with HQC matching coupled-cluster (CCSD) accuracy in fragment energies.
{We present the first HQC protein-ligand binding prediction using a mixed-basis set and an automated end-to-end workflow enabling practical HQC calculations of large protein systems.}
\end{abstract}

\section{Description}

{Largest, most resource-intensive} heterogeneous quantum-classical (HQC) electronic-structure calculation to date -- 12,635 atoms (31,795 orbitals), $210\times$ in accuracy and $>40\times$ in size beyond state-of-the-art. Two 156-qubit quantum processors, using up to 94 qubits, running over {239} hours, executing {21,006} circuits and collecting {$\scinot{3.0}{9}$} samples. {Automated end-to-end workflow, systematically improvable accuracy.}

\begin{table}[h!]
\centering
\caption{Summary of performance attributes}
\setlength{\tabcolsep}{2pt}
\begin{tabular}{ll}
\toprule
\textbf{Performance attribute} & \textbf{This submission} \\
\midrule
Category of Achievement & Scalability, Peak Performance \\
& Time to solution \\
Type of method used & Implicit: EWF-TrimSQD embedding with\\ 
& heterogeneous quantum-classical solver \\
Results reported on the basis of & Whole application including I/O \\
Precision reported & Double precision \\
System scale & Results measured on full-scale system \\
Measurement mechanism & Timers\\
\bottomrule
\end{tabular}
\label{tab:performance}
\end{table}

\section{Overview of the problem}

Modeling biomolecular systems accurately is a grand challenge in contemporary biochemistry, with important implications in areas including structure-based drug discovery. Because experimental screening and characterization of drug-protein complexes are costly and time-consuming, accurate computational guidance has the potential to reduce the time and expense of the drug discovery pipeline, making this a technologically and socially impactful research area in high-performance computing (HPC). However, quantitative predictions require chemical accuracy, i.e., kcal/mol agreement between experimental and computed energy differences, e.g. protein-ligand binding energies, combined with scalability to biologically relevant system sizes of thousands of atoms.

The primary challenge towards this goal is to solve the computationally costly quantum-mechanical many-electron Schr\"{o}dinger equation (SE) with accurate and scalable methods. This task typically involves density functional theory (DFT) or post-Hartree-Fock wavefunction (WF) methods. DFT offers more scalable solutions~\cite{ching2021ultra}, but is based on semiempirical approximations that may not produce accurate results and lead to false positives and false negatives, limiting its usefulness in drug discovery campaigns. WF methods may provide systematically improvable approximations, which is critical for adequately modeling biomolecular energetics and functionalities, but are too resource-intensive for application to protein-sized systems (see Sec.~\ref{sec:sota}).

An open question in this field is whether and how quantum computation can support the in-silico investigation of biochemically relevant systems.  Current pre-fault-tolerant and projected early fault-tolerant quantum processors (QPUs) and the algorithms executing on these platforms can deliver systematically improvable SE solutions with a favorable tradeoff between accuracy and computational cost~\cite{yoshioka2024hunting,goings2022reliably}. However, the natures of both categories of devices and algorithms are such that solutions can be delivered for chemical systems that are orders of magnitude smaller than those encountered in biomolecular applications, e.g. protein-ligand complexes.

Importantly, challenges of comparable scale and difficulty arise beyond biochemistry, in fields such as heterogeneous catalysis, energy materials, and electronics. Across these domains, the gap between systems relevant for technological applications and those accessible by WF methods has prompted the development of quantum embedding methods~\cite{stocks2024breaking, Barca2022FMOMP2, nakai2023divide, fedorov2023complete, broderick2025fragme}. These methods (see Sec.~\ref{sec:sota}) partition the system into multiple smaller subsystems or ``fragments'' that can be treated independently computationally and recombined to access properties of the system. 

By exploiting spatial locality and exposing coarse-grained parallelism, quantum embedding methods offer a promising path to scalable quantum-mechanical simulations of biomolecules. In this context, QPUs can operate alongside CPUs and GPUs as specialized co-processors. Within such heterogeneous HPC environments, quantum algorithms can be invoked as specialized subroutines to accurately solve the SE for specific fragments in the context of a larger classical workflow~\cite{lim2024fragment,ma2023multiscale,vorwerk2022quantum}. In this way, quantum processors and algorithms complement their classical counterparts rather than serving as standalone replacements.

Although quantum embedding naturally provides a structural foundation for heterogeneous quantum-classical (HQC) workflows, two challenges prevent its application to systems of biomolecular scale with sufficient accuracy and within realistic time scales: (i) each fragment must be represented by a compact yet accurate basis of electronic states, whose construction may require full-system WF calculations at substantial cost; and (ii) the fragment SE requires solutions of systematically improvable accuracy and favorable accuracy-cost tradeoff. Both challenges place significant pressure on memory bandwidth, accelerator usage, and inter-node communication, requiring algorithmic advances and software stack improvements to enable broader applicability and scalability.

\begin{figure*}[tb]
  \centering
  \includegraphics[width=0.95\linewidth]{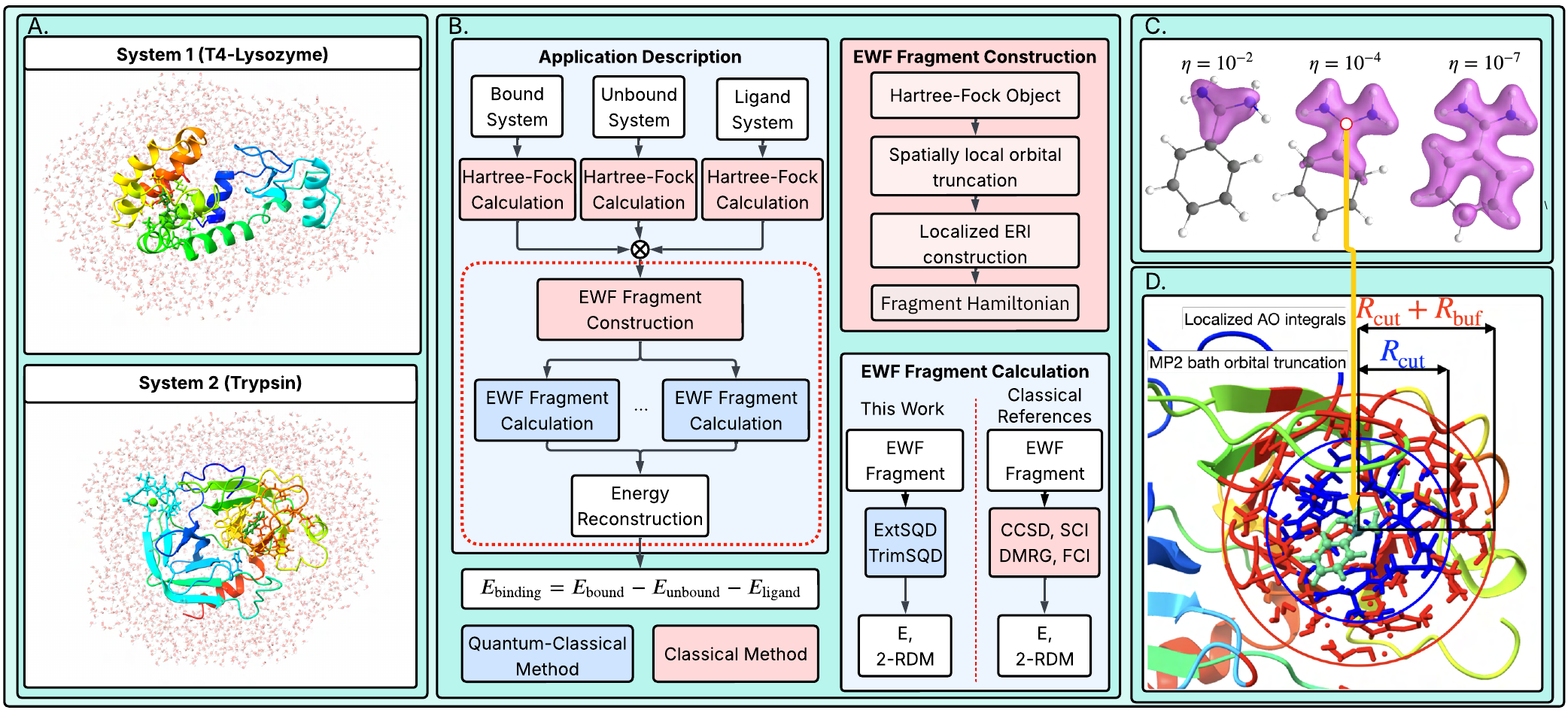}
  \caption{Systems investigated (A), overarching algorithmic scheme adopted, (B), representation of a fragment (C), and details of the lower-scaling fragment construction (D) introduced in this work. {We performed workflow automation on computational tasks in panel B (red dashed frame).}
}
  \label{fig:workflow}
\end{figure*}

{\em{Summary of contributions.}} This work presents a substantial leap in modeling molecules with quantum embedding and HQC fragment solvers, {culminating in the simulation of two protein-ligand complexes of unprecedented size (Fig.~\ref{fig:workflow}a) and the first prediction of protein-ligand binding using this approach.} 
Its main algorithmic innovations, Sec.~\ref{sec:innovations}, are: (i) a quantum embedding method (EWF) with $O(1)$-scaling and embarrassingly parallel fragment construction; (ii) TrimSQD, an HQC method for fragment calculations that delivers an improvement of up to 210$\times$ in accuracy over the state-of-the-art ExtSQD, Sect.~\ref{sec:sota}; and (iii) GPU-accelerated Selected-Basis Diagonalization (SBD-G), a memory-efficient linear algebra kernel that allows scalable execution across GPU nodes. As shown in Table~\ref{table:watermark}, these innovations allow the most resource-intensive HQC computation for quantum chemistry to date, at 12,000-atom scale ($>40\times$ larger than state-of-the-art), leveraging two 156-qubit quantum processors (using up to 94 qubits, operating over {239} hours, executing {21,006} circuits, and collecting ${\scinot{3.0}{9}}$ measurement outcomes) alongside {three supercomputers (Fugaku, Miyabi-G, and ROQUO)}, achieving 72.5\% parallel efficiency.

\section{Current state of the art}
\label{sec:sota}

We provide an overview of significant developments in quantum embedding calculations with HQC fragment solvers. A subset of these developments is summarized in Table~\ref{table:watermark}.

{\em{Schr\"{o}dinger equation.}} The quantum-mechanical ground state of a system of electrons is the lowest-energy solution of the time-independent many-electron Schr\"{o}dinger equation. In a finite basis it takes the form of an eigenvalue equation,
\begin{equation}
\label{eq:schrodinger}
\sum_{\bts{y}} H_{\bts{x} \bts{y}} \, \Psi_{\bts{y}} 
= 
E \, \Psi_{\bts{x}}
\;,
\end{equation}
where the vector of components $\Psi_{\bts{x}}$ represents the ground state, the matrix of elements $H_{\bts{x} \bts{y}}$ represents the quantum-mechanical energy operator or ``Hamiltonian'', its lowest eigenvalue $E$ is the ground-state energy of the system, and $\bts{x} \in \{0,1\}^{2M}$ labels an ``electronic configuration'', i.e. a pattern of occupation of $M$ spatial orbitals ($o$) by $N$ electrons ($e$). Since the number of configurations increases as $\binom{M}{N}$, exact solutions of Eq.~\eqref{eq:schrodinger} provided by the full configuration interaction method (FCI) are restricted to small and chemically unrealistic contexts. Consequently, practical methods introduce approximations.

{\em{Classical WF methods.}} WF methods offer a first-principles approach to quantum chemistry by approximately solving Eq.~\eqref{eq:schrodinger} with algorithms running on classical HPC in isolation or in synergy with QPUs. Some WF methods for classical HPC with systematically improvable accuracy, required to adequately describe complex biochemical systems, are
\begin{itemize}
\item Density matrix renormalization group (DMRG): approximates the solution of Eq.~\eqref{eq:schrodinger} by variationally optimizing a tensor network ansatz. A bond dimension parameter controls the maximum quantum-mechanical correlations (``entanglement'') the network can represent, and with it the accuracy and computational cost of the method. The largest study to date is for a system of \actspace{89}{102}~\cite{2026largedmrg}, an order of magnitude smaller than the Trp-cage mini-protein in Table~\ref{table:watermark}.
\item Selected configuration interaction (SCI) methods: approximate $\Psi_{\bts{x}}$ with a sparse vector supported on a subset of selected configurations $\bts{x}$~\cite{holmes2016heat}. The accuracy of SCI is systematically improvable by increasing the number of selected configurations, with different flavors offering specific accuracy-cost tradeoffs based on their selection criteria and implementation details. The largest study to date, from TrimCI~\cite{zhang2025from}, is for a system of \actspace{114}{73}, an order of magnitude smaller than the Trp-cage mini-protein in Table~\ref{table:watermark}.
\end{itemize}

{\em{Sample-based quantum diagonalization (SQD).}} As listed in Table~\ref{table:watermark}, the ``high watermark'' for heterogeneous quantum-classical simulations of molecules in isolation (i.e. not within quantum embedding) is for \actspace{54}{36} and \actspace{46}{50} systems, an order of magnitude smaller than the Trp-cage mini-protein in Table~\ref{table:watermark}. These computations were made possible by a family of methods {including SQD~\cite{robledo2025chemistry,barison2025quantum,shirakawa2025closed} and conceptually related QSCI~\cite{kanno2023quantum}}, which can be understood as HQC counterparts of SCI. Within SQD, QPUs sample configurations $\bts{x}$ by measuring one or more quantum circuits (providing a configuration selection criterion), and CPUs/GPUs mitigate errors in the sampling step and solve the projection of Eq.~\eqref{eq:schrodinger} in the subspace of selected and error-mitigated configurations. The accuracy of SQD is systematically improvable by increasing the number of selected configurations and the criterion to select configurations (e.g., standard SQD \cite{robledo2025chemistry} and SQDrift~\cite{piccinelli2026note} use parametrized ansatz-based and random time-evolution quantum circuits to draw samples, respectively, and ExtSQD extends the set of configurations $\bts{x}$ by including a subset of $\bts{y}$ such that $H_{\bts{x}\bts{y}} \neq 0$~\cite{barison2025quantum}). Due to its noise resilience and synergy with CPUs/GPUs, SQD can be executed on existing pre-fault-tolerant QPUs, while remaining useful when projected fault-tolerant QPUs will become available (e.g., as a starting point for other algorithms like Krylov quantum diagonalization \cite{yoshioka2025krylov} and quantum phase estimation \cite{yoshioka2024hunting}).

\begin{table*}[t!]
\centering
\caption{Largest ground-state energy calculations from heterogeneous quantum-classical methods, in isolation and within quantum embedding. Systems investigated in this work are shown in Fig.~\ref{fig:workflow}a }~\label{table:watermark}
\setlength{\tabcolsep}{3.5pt}
\begin{tabular}{cccccccc}
\toprule
\header{Method} & \header{Benchmark system} & \header{Atoms} & \header{Active space} & \header{Embedding} & \header{Features} & \header{HQC error}$^{\ddagger}$ & \header{Ref.} \\
 &  &  &  & \header{(largest {HQC})} & & \header{(largest {HQC})} &  \\
\midrule
SQD & [4Fe-4S] & 28 & \actspace{54}{36} & N/A & 1. ansatz-based circuit & 250 mHa$^{*}$ & \cite{shirakawa2025closed} \\
  &   &   &   &   & 2. closed-loop parametrization & \\
  &   &   &   &   & 3. configuration carryover & \\
  &   &   &   &   & 4. Selected basis diagonalization & \\
\midrule
ExtSQDrift & $\mathrm{C_{13}Cl_{12}}$ & 15 & \actspace{46}{50} & N/A & 1. random time-evolution circuits & 75 mHa & \cite{piccinelli2026note} \\
 &  &   &   &   & 2. ExtSQD &  \\
\midrule
EWF-ExtSQD & Trp-cage & 303 & \actspace{1,158}{919} & EWF & 1. ansatz-based circuit & 21 mHa$^\dagger$ & \cite{shajan2026quantumprotein} \\
  &  &  &  & \actspace{34}{33} & 2. classical parametrization &  \\
  &  &  &  &   & 3. configuration carryover &  \\
  &  &  &  &   & 4. ExtSQD &  \\
\toprule
EWF-TrimSQD & trypsin & 12,635 & (43,858e, 31,795o) & EWF + Sec.~\ref{sec:innovations} & Sec.~\ref{sec:innovations} & 0.10 mHa$^\dagger$ & this work \\
{STO-3G basis} & benzamidine &  &  & \actspace{66}{45} & &  \\
  \midrule
EWF-TrimSQD & T4-Lysozyme & 11,608 & \actspace{39,880}{28,844} & EWF + Sec.~\ref{sec:innovations} & Sec.~\ref{sec:innovations} & 0.16 mHa$^\dagger$ & this work \\
{STO-3G basis}  & $n$-butyl-benzene & & & \actspace{36}{35} & &  \\
  \midrule
EWF-TrimSQD & T4-Lysozyme & 11,608 & {\actspace{39,880}{29,455}} & EWF + Sec.~\ref{sec:innovations} & Sec.~\ref{sec:innovations} & {2.23} mHa$^\dagger$ & this work \\
{mixed basis$^\S$}  & $n$-butyl-benzene & & & {\actspace{34}{42}} & &  \\
\bottomrule
\end{tabular}
\\
\scriptsize{$^{\ddagger}$ absolute deviation between HQC energy and reference classical DMRG energy, $^{*}$ milliHartree (atomic) units, $^\dagger$ in the largest EWF fragment calculation,
$^\S$ mixed basis denotes the use of 3-21G orbitals on the ligand and nearby residues, and STO-3G orbitals on the remainder of the system
}
\end{table*}

{\em{Embedded wavefunction (EWF).}} WF methods have a natural and compelling application as fragment solvers in quantum embedding calculations. As listed in Table~\ref{table:watermark}, the ``high watermark'' for molecular sizes achieved in quantum embedding calculations with heterogeneous quantum-classical fragment solvers is the 303-atom Trp-cage mini-protein. This study was made possible using the embedded wavefunction (EWF) formalism \cite{nusspickel2023effective}. EWF exploits the fundamental locality of entanglement in biomolecular systems to partition a molecule into ``fragments''. As shown in Fig.~\ref{fig:workflow}b, each fragment is solved independently with a high-level WF method while its surrounding environment is represented in a simplified but systematically improvable way. The fidelity of this environment is controlled by a single parameter $\eta$ called the ``bath truncation threshold''. As shown in Fig.~\ref{fig:workflow}c, each fragment comprises an atom and a set of electrons and orbitals (bath) surrounding the atom and entangled with it: as $\eta$ decreases, the fragment increases in size and, for $\eta\to0$, it converges to the full molecule. Correspondingly, EWF converges to the full-system description provided by the high-level WF method. Furthermore, the EWF fragments are approximately independent, which exposes more opportunities for parallelizing their solutions. Although the landscape of quantum embedding methods is broad and diverse~\cite{lim2024fragment,ma2023multiscale,vorwerk2022quantum}, here we focus on EWF as other methods have enabled HQC studies at the scale of drug-like molecules (e.g., amantadine~\cite{patra2026towards}), an order of magnitude smaller than the Trp-cage mini-protein in Table~\ref{table:watermark}.

\section{Innovations realized}
\label{sec:innovations}

{\em{Limitations of EWF.}} While EWF is rigorous and systematically improvable, its application at the biomolecular scale presents implementation challenges, listed in Table~\ref{tab:innovations}. Fragment construction requires evaluation of two-electron repulsion integrals {(ERI) on the full system, and a perturbation theory (MP2) calculation, with overall $O(M^3)$ memory, runtime, and I/O requirements}. The fragment solution is challenged because: (i) SQD and ExtSQD may yield inaccurate results, see Table~\ref{table:watermark}, and (ii) since the number of EWF fragments grows as $O(M)$, calculations at biochemical scale demand a diagonalization kernel optimized for GPU-accelerated, distributed-memory execution across thousands of fragments.

{\em{Summary of innovations and impact.}} In summary, the current state of the art in EWF calculations with HQC fragment solvers is limited by either (i) the cost with respect to system size for the quantum embedding or (ii) the accuracy of the fragment solver. This study {improves scalability and accuracy through (i) a lower-scaling formulation of EWF, (ii) higher-accuracy TrimSQD fragment solutions, (iii) an automated end-to-end workflow}, and does so through the innovations listed in Table~\ref{tab:innovations}.

As shown in Table~\ref{table:watermark}, when these innovations are implemented in an optimal fashion, HQC calculations based on quantum embedding and TrimSQD fragment solutions are possible at 12,000-atom scale, $>40\times$ the previous state of the art in HQC (i.e. 12,635 versus 303 atoms). In addition, HQC fragment solutions have substantially higher accuracy than previous studies based on SQD and ExtSQD: e.g., the deviation between HQC and DMRG energies is 21 mHa for the largest Trp-cage fragment~\cite{shajan2026quantumprotein} and 0.1 mHa for the largest trypsin fragment (i.e. $210\times$ accuracy improvement).

\begin{table*}[t!]
  \centering
  \renewcommand{\arraystretch}{1.2}
  \caption{Summary of limitations in conventional EWF and the corresponding
           innovations introduced in this work.}
  \label{tab:innovations}
  \begin{tabular}{>{\raggedright\arraybackslash}p{0.54\textwidth} >{\raggedright\arraybackslash}p{0.34\textwidth} c}
    \toprule
    \textbf{Limitation} & \textbf{Innovation} & \textbf{Section} \\
    \midrule

    Bath construction requires {$O(M^3)$} in memory and runtime
      & Spatially localized orbital truncation
      & \ref{sec:slot} \\

    ERI evaluation requires global four-center AO integrals; {$O(M^3)$} memory and I/O
      & Localized ERI construction
      & \ref{sec:leri} \\

    ExtSQD diagonalizes subspaces independently, limiting combined subspace quality
      & TrimSQD
      & \ref{sec:trimsqd} \\
      
    Subspace diagonalization dominates the computational cost in SQD
      & GPU-accelerated Selected-Basis Diagonalization 
      & \ref{sec:SBD} \\
      
    Manually {coordinated workflow spanning multiple computational resources} 
      & Automated {end-to-end workflow}
      & \ref{sec:workflow-automation} \\
    \toprule
  \end{tabular}
\end{table*}

\subsection{Overarching algorithm}
\label{sec:overarching}

The algorithm, shown in Fig.~\ref{fig:workflow}b, consists of these steps:
\begin{enumerate}
\item Given the geometry of a molecule we perform a Hartree-Fock calculation, at cost $O(M^3)$ where $M$ is the number of orbitals in the full system, to generate a set of molecular orbitals that are the input of quantum embedding.
\item We decompose the system into fragments. In this ``EWF fragment construction'' step we introduce two innovations, detailed in Sec.~\ref{sec:slot} and Sec.~\ref{sec:leri}, that reduce the scaling from {$O(M^3)$} to $O(1)$ per fragment and achieve embarrassing parallelization across fragments. We use $\eta = 10^{-5}$ in the protein and ligand region and $\eta = 10^{-7}$ in the surrounding H$_2$O molecules.
\item For each EWF fragment $c$, with spatial orbitals $M_c \ll M$, we solve Eq.~\eqref{eq:schrodinger} in parallel, with:
\begin{itemize}
\item FCI for fragments centered on H atoms, with $M_c < 13$
\item TrimSQD (see Sec.~\ref{sec:trimsqd}) for fragments centered on heavier atoms, with {$13 \leq M_c \leq M_c^{*}$ where $M_c^*$ is a user-defined maximum fragment size  compatible with error rates of employed QPUs}
\item CCSD (coupled-cluster singles and doubles), a widely used WF method for classical HPC, to provide a point of comparison for TrimSQD {and to simulate fragments with $M_c > M_c^*$, if any}
\end{itemize}
\item The EWF fragment calculations by ExtSQD and TrimSQD, as shown in Fig.~\ref{fig:TrimSQD-flow}, comprise a ``quantum sampling'' step, a ``subspace diagonalization'' step, and a ``subspace extension'' step. In the quantum sampling step, for each fragment we construct a parametrized ansatz-based quantum circuit~\cite{motta2023bridging,lin2025improvedparameterinitializationlocal}. We execute it on a QPU and measure it $S$ times, collecting measurement results $\{ \bts{x}_k \}_{k=1}^S$ that correspond to electronic configurations, used to guide the subspace diagonalization. In this study, we use two QPUs, see Sec.~\ref{sec:performance}.
In the ``subspace diagonalization'' step, we introduce two innovations detailed in Sec.~\ref{sec:trimsqd} and Sec.~\ref{sec:SBD}, that substantially improve the accuracy of heterogeneous quantum-classical fragment solutions.
\item In the final step ``Energy reconstruction'', we {estimate} the total energy {from the partitioned one-body density matrix and two-body cumulant of the fragment solutions~\cite{nusspickel2023effective}. We evaluate one- and two-body contributions independently for each fragment,} in parallel, using strictly local quantities to avoid memory and runtime bottlenecks.
\end{enumerate}

\subsection{Innovations in EWF fragment construction}

The two key innovations that eliminate data dependencies at full-system scale and render the workflow embarrassingly parallel across fragments are the following:

\subsubsection{Spatially localized orbital truncation}
\label{sec:slot}

The standard construction of EWF fragments requires {ERI generation and MP2 calculation. At $M=31,795$ orbital scale, this is computationally intractable}.
We lift this limitation by restricting, for each fragment centered around an atom at position ${\bf{R}}$, the MP2 calculation to molecular orbitals that are spatially localized within a sphere of center ${\bf{R}}$ and radius $R_{\text{cut}} = 7~\text{\AA}$\footnote{benchmarked on the Trp-cage mini-protein across $R_{\text{cut}} = 3,5,7,10~\text{\AA}$}, as shown in Fig.~\ref{fig:workflow}d.
This choice is empirical, but grounded in the notion that entanglement is local in typical biomolecular systems: when quantum-mechanical correlations decay rapidly with distance, the molecular orbitals that govern the entanglement between an atom and its environment are predominantly localized in the geometric neighborhood of the atom.
If $R_{\text{cut}}$ is sufficiently large, we can capture physically relevant contributions to the environment while discarding negligibly small long-range terms.

\subsubsection{Localized electron repulsion integral construction}
\label{sec:leri}

Even after orbital truncation, a latent bottleneck remains in conventional EWF implementations: the ERIs required for the MP2 calculation are typically obtained by transforming full-system four-center {atomic-orbital (AO)} integrals into the basis of fragment molecular orbitals.
For large systems, the construction and transformation of these integrals {using density fitting imposes $O(M^3)$} scaling with a large prefactor and {$O(M^3)$} memory and I/O requirements.

We eliminate this bottleneck by restricting the construction of ERIs to molecular orbitals localized in a region of center ${\bf{R}}$ and radius $R_{\text{cut}}+R_{\text{buf}}$ with $R_{\text{buf}} = 3~\text{\AA}$, as shown in Fig.~\ref{fig:workflow}d.
Molecular orbitals outside this sphere contribute negligibly to the ERIs within the fragment, since the overlap between localized functions decays exponentially with the distance between their centers. 
The choice $R_{\text{buf}} = 3~\text{\AA}$ is empirical and balances computational cost with minimization of artificial boundary effects at the edge of the MP2-treated region.

The combination of spatially localized orbital truncation and localized electron repulsion integral construction reduces the cost of EWF fragment construction from {$O(M^3)$} to $O(1)$ per fragment and renders the operation embarrassingly parallel.

\begin{figure}[h!]
  \centering
  \includegraphics[width=\columnwidth]{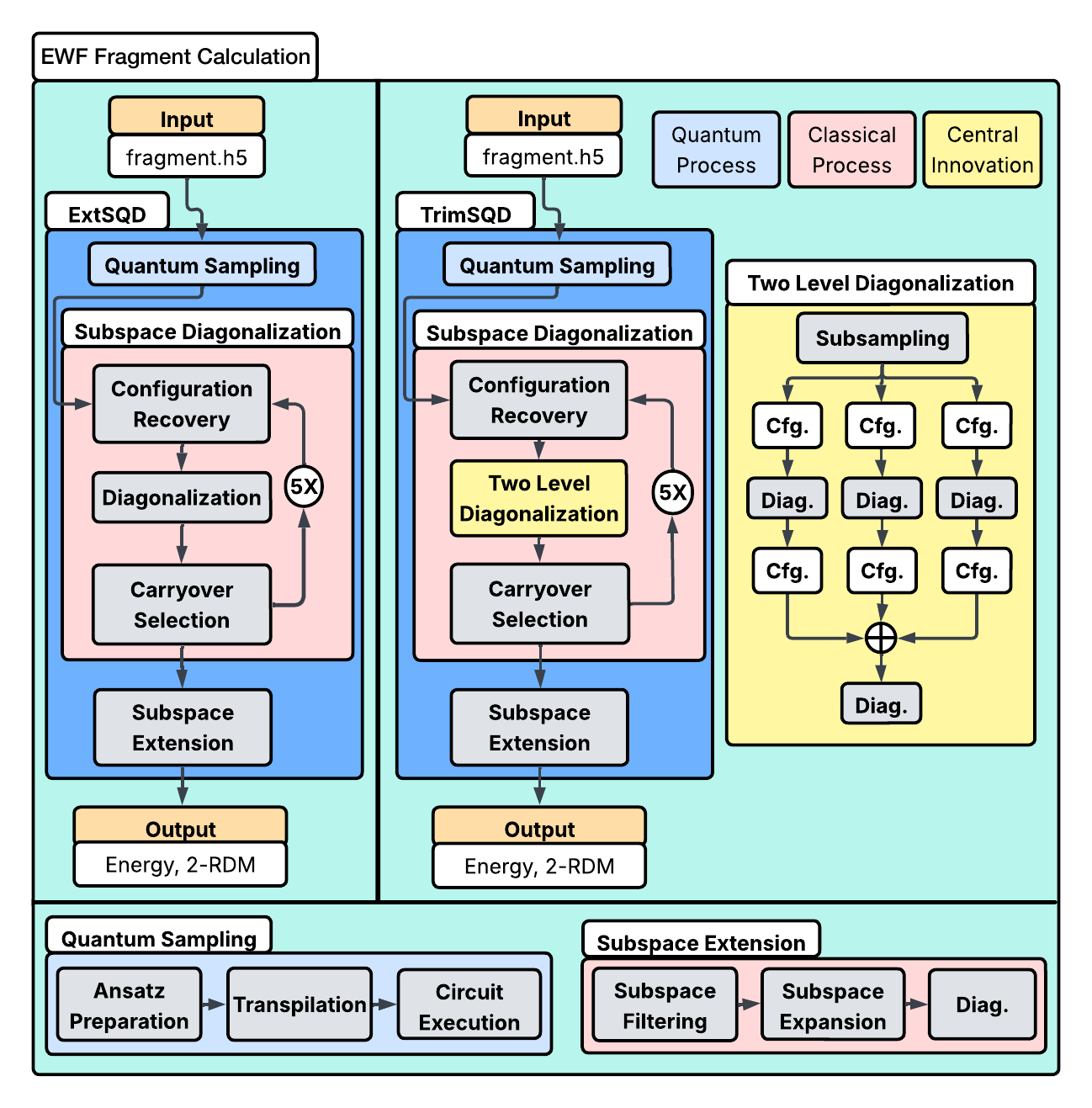}
  \caption{The workflow of ExtSQD (left) and TrimSQD (middle), with central innovations from this study marked in light yellow (right). ``Cfg.'' is an abbreviation for ``electronic configurations'', $\bts{x}$ {in Eq.~\eqref{eq:schrodinger} of} the main text.}
  \label{fig:TrimSQD-flow}
\end{figure}

\subsection{Innovations in EWF fragment calculations}

Fig.~\ref{fig:TrimSQD-flow} (left) shows the workflow of state-of-the-art ExtSQD: after quantum sampling (light blue), we (i) apply an iterative self-consistent configuration recovery~\cite{robledo2025chemistry} to mitigate errors in noisy quantum samples and produce an improved set of configurations; (ii) solve the projection of Eq.~\eqref{eq:schrodinger} onto the subspace of recovered configurations; (iii) carry over the configurations contributing most to the solutions of Eq.~\eqref{eq:schrodinger} to the next iteration of (i). Finally, we perform a subspace extension (bottom right pink block in Fig.~\ref{fig:TrimSQD-flow}).

\subsubsection{TrimSQD}
\label{sec:trimsqd}

To achieve higher accuracy through improved configuration selection, in this study we introduce TrimSQD. This new variant incorporates the idea of trimming determinants based on diagonalization results, as in closed-loop SQD~\cite{shirakawa2025closed} and TrimCI~\cite{zhang2025from}, into ExtSQD.
The TrimSQD workflow is shown in Fig.~\ref{fig:TrimSQD-flow} (middle) and consists of:

\begin{enumerate}[label=\Roman*.]
  \item quantum sampling (light blue block in Fig.~\ref{fig:TrimSQD-flow}, see~\ref{sec:overarching})
  \item configuration recovery loop:\label{cr-step}
  \begin{enumerate}[label=\roman*)]
    \item we apply a self-consistent configuration recovery~\cite{robledo2025chemistry}
    \item we merge subsampled configurations with those carried over from the previous configuration recovery iteration
    \item we divide configurations and MPI ranks into subgroups. For each subgroup, we perform diagonalization on assigned configurations and carry over to the next step the top $k_1 \%$ configurations by contribution to $|\Psi_{\bts{x}}|$
    \item we collect configurations carried over from all subgroups, perform subspace diagonalization across all MPI ranks, and retain the top $k_2 \%$ configurations by contribution~\label{key-step}
  \end{enumerate}
  \item we extend the subspace of configurations $\bts{x}$ by including all $\bts{y}$ with $H_{\bts{x}\bts{y}} \neq 0$ and the Hamming distance from $\bts{x}$ at most two~\cite{barison2025quantum} (bottom-right pink block in Fig.~\ref{fig:TrimSQD-flow}).
  \item we perform diagonalization over the extended subspace on all MPI ranks. For the final diagonalization, we use configurations such that $|\Psi_{\bts{x}}| > \varepsilon$, where $\varepsilon$ is a user-defined threshold. Compared to the previous step, the dimension of the expanded subspace becomes very large, so this phase typically dominates the execution time.
\end{enumerate}

\subsubsection{Selected-basis diagonalization}
\label{sec:SBD}

Subspace diagonalization is the dominant contribution to the computational cost of ExtSQD and TrimSQD, and requires an implementation tailored for large-scale parallel execution.
In this work, we employ our in-house diagonalization code, Selected-Basis Diagonalization (SBD)~\cite{sbd_repository}. 
SBD adopts a distributed-memory design in which a single vector $\Psi_{\bts{x}}$ is partitioned and stored between multiple computing nodes, thereby reducing the memory footprint per node and allowing scalable computations for large problem sizes. 
We achieve additional performance improvements through careful code optimization and GPU acceleration~\cite{doi2026gpu}. 
In particular, the matrix-vector multiplication kernel, which dominates the computational cost in subspace diagonalization, is redesigned using fine-grained data-parallel primitives and flattened GPU-resident data layouts to efficiently exploit massive thread-level parallelism.
The implementation leverages portable data-parallel programming abstractions for GPU execution, enabling efficient use of modern accelerator architectures.
To improve parallel efficiency, nested excitation loops are restructured to expose a high degree of concurrency, while accumulation conflicts are resolved through atomic operations.
In contrast to prior approaches that rely on explicit configuration generation and its caching, we eliminate configuration generation altogether by introducing auxiliary helper data structures that directly encode excitation mappings, thereby removing a major computational bottleneck and reducing both memory overhead and computational cost.
These optimizations collectively enable high-performance scalable diagonalization on GPU-accelerated HPC systems.

Step~\ref{cr-step}-\ref{key-step} is the central innovation that distinguishes TrimSQD from ExtSQD (yellow block in Fig.~\ref{fig:TrimSQD-flow}). It improves configuration selection by introducing interactions between subgroups of otherwise independent configurations. In conjunction with optimized GPU-accelerated distributed-memory execution, TrimSQD delivers more accurate fragment energies.

{\subsection{Workflow automation}
\label{sec:workflow-automation}}

{Beyond algorithmic innovations, we developed a Prefect-based~\cite{prefect} end-to-end orchestration framework for automated execution across QPUs and GPU/CPU-based HPC systems.
A persistent workflow controller service on a login node coordinates jobs submitted through the native HPC scheduler with quantum circuits (jobs) submitted through IBM Quantum Runtime to multiple backends, presenting the full heterogeneous computation across these systems as a single monitored workflow. The framework orchestrates fragment construction, quantum sampling, heterogeneous fragment-solver execution, and energy reconstruction (see Fig.~\ref{fig:workflow}b).

The workflow advances asynchronously at fragment granularity without requiring any global barrier between systems. Fragments are dynamically routed through heterogeneous computational paths, selectively invoking FCI, CCSD, or TrimSQD based on fragment size and accuracy requirements. Fragments routed to TrimSQD undergo quantum sampling on QPUs followed by GPU-accelerated subspace diagonalization on the HPC system.
Workflow controllers enforce concurrency limits, reconcile persistent job registries with both the HPC schedulers and IBM Quantum Runtime, and submit additional work as capacity becomes available.

Job identifiers, execution state, and validated output markers persist across controller restarts, enabling reuse of completed work and selective resubmission of failed tasks without re-executing the full pipeline. 
This pipelined, multi-backend, and restartable execution model enabled the sustained heterogeneous computation reported in Sec.~\ref{sec:TTS} while substantially reducing manual coordination.}

\vspace{6pt}
\section{How performance was measured}
\label{sec:performance}

\subsection{Benchmark molecular systems}
We selected two protein–ligand complexes as benchmarks, to represent contrasting binding mechanisms, and to assess the portability of our workflow. The sizes of these complexes are reported in Table~\ref{table:watermark}.

For each complex, we obtain an initial geometry from a PDB crystal structure and perform 3 steps of classical molecular dynamics followed by NVT (Number of particles, Volume, Temperature) and NPT (Number of particles, Pressure, Temperature) equilibration with the AMBER ff14SB force field and the AMBER26 package~\cite{case2025recent}. We assign protonation states at $\mathrm{pH}\approx 7$ using the H++ server~\cite{gordon33h++}, then carefully inspect structures to deprotonate selected Histidine and Lysine residues while retaining charge neutrality of the entire system and leaving active sites unaffected. We compute partial charges for ligands using the AM1-BCC method~\cite{jakalian2002fast}. We perform all quantum-mechanical calculations with the STO-3G minimal basis set, consistent with the state-of-the-art Trp-cage benchmark~\cite{shajan2026quantumprotein}.

\subsubsection{trypsin, benzamidine {(system I)}}

trypsin is a type of serine protease enzyme, and benzamidine is its well-known inhibitor. The trypsin-benzamidine complex is a well-established benchmark for the development of new methodologies in protein-ligand interaction studies of hydrophilic protein cavities~\cite{buch2011complete}. The complex comprises 12,635 atoms and 31,795 molecular orbitals {in the STO-3G basis}, with 3,135 water molecules retained within a solvent cutoff of 10~\AA{} from the protein boundary. We import crystal coordinates from PDB ID: 3PTB~\cite{marquart1983geometry}.

\subsubsection{T4-Lysozyme, $\textit{n}$-butyl-benzene {(system II)}}

T4-Lysozyme is an engineered protein widely used for benchmark studies of hydrophobic interactions in protein-ligand systems, while $\textit{n}$-butyl-benzene is a ligand with high affinity to the T4-Lysozyme cavity~\cite{merski2015homologous, morton1995energetic}. The complex comprises 11,608 atoms and {28,844 (29,455) molecular orbitals in the STO-3G (mixed) basis}, with 2,986 water molecules retained within a solvent cutoff of 10~\AA{} from the protein boundary.  We import crystal coordinates from PDB ID: 4W57~\cite{merski2015homologous}.

\subsection{{Summary of calculations}}

{We carried out six calculations, summarized in Table~\ref{tab:summary}, and organized into: 
(i) ``baseline'' calculations (A,B) yielding the binding energies of systems I and II in a minimal STO-3G basis, 
(ii) ``portability'' calculations (C,D) identical to A,B with diagonalization carried out on a different HPC platform; 
(iii) a ``refinement'' calculation yielding the binding energy of system II in a mixed basis (more realistic than STO-3G);
(iv) a ``time-to-solution'' calculation using the automated end-to-end workflow to yield the ground-state energy of system II.
}

\begin{table}[h!]
\centering
\caption{Summary of calculations}
\label{tab:summary}
\setlength{\tabcolsep}{1.5pt}
\begin{tabular}{ccccccc}
\toprule
{\header{Label}} & {\header{Category}} & {\header{System}} & {\header{Basis}} & {\header{Target}} & {\header{QPUs}} & {\header{HPC}} \\ \midrule A & baseline & I & STO-3G & BE$^\S$ & \device{c} & Fugaku \\ B & baseline & II & STO-3G & BE & \device{c,k} & Fugaku \\ \midrule C & portability & I & STO-3G & BE & ---$^\dagger$ & Miyabi-G \\ D & portability & II & STO-3G & BE & ---$^\ddagger$ & Miyabi-G \\ \midrule E & refinement & II & mixed & BE & \device{c,k} & Fugaku \\ \midrule F & time to solution & II & mixed & TTS$^\P$ & \device{c,k} & ROQUO \\
\bottomrule
\end{tabular}
\\
\scriptsize{
$^\S$ binding energy
$^\P$ time to solution
$^\dagger$ same samples as A 
$^\ddagger$ same samples as B
}
\end{table}

\subsection{HPC and Quantum computing platforms}

\subsubsection{Quantum sampling}

We distribute quantum sampling across two 156-qubit Heron r2 QPUs, \device{cleveland} and \device{kobe} (henceforth abbreviated as \device{c} and \device{k}). We allocate samples based on fragment size as detailed in Table~\ref{tab:QPUs} and report the largest quantum circuits executed in Table~\ref{tab:atoms-for-qpu}.

\begin{table}[h!]
\centering
\caption{Distribution of quantum sampling across QPUs {for binding energy calculations A, B, E in Table~\ref{tab:summary}}.}
\label{tab:QPUs}
\setlength{\tabcolsep}{1.5pt}
\begin{tabular}{cccrccrc}
\toprule
\textbf{Calc.}&\textbf{Frag.} & \textbf{MO} & \textbf{Number} & \textbf{$S$} & \textbf{QPUs} & \textbf{QPU time} & \textbf{Complete}\\
 & \textbf{group} & & \textbf{of frag.} & & & \textbf{per frag.} &\textbf{QPU time}\\
\midrule
{A} & 1 & 13-26 & 4,296 & 100,000 & \device{c} & $\approx$ 29s & 56 h\\
 & 2 & 27-45 & 549 & 500,000 & \device{c} & $\approx$ 2m22s &\\
\midrule
{B} & 3& 13-18 & 751 & 100,000 & \device{k} & $\approx$ 29s & 50 h\\
        & 4& 19-26 & 3,228 & 100,000 & \device{c}  & $\approx$ 29s &\\
        & 5& 27-35 & 467 & 500,000 & \device{c} & $\approx$ 2m22s& \\
\midrule
{E} & 6& 13-18 & 374 & 25,000 & \device{c,k} & $\approx$ 7s & 83 h\\
  & 7& 19-26 & 5164 & 100,000 & \device{c,k} & $\approx$ 29s &\\
        & 8& 27-35 & 1697 & 250,000 & \device{c,k} & $\approx$ 1m11s& \\
        & 3-21G& 28-42 & 184 & 500,000 & \device{c,k} & $\approx$ 2m22s& \\
\bottomrule
\end{tabular}
\end{table}

\begin{table}[h!]
\centering
\caption{Benchmark circuits used for QPU performance analysis {for binding energy calculations A, B, E in Table~\ref{tab:summary}}.}
\label{tab:atoms-for-qpu}
\setlength{\tabcolsep}{1.5pt}
\begin{tabular}{cccccc}
\toprule
\multirow{2}{*}{\header{Calc.}}
& \multirow{2}{*}{\header{Fragment} \,} 
& \multicolumn{4}{c}{\header{Quantum resource count}}\\
\cline{3-6}
& & 1-qubit gates & 2-qubit gates & 2-qubit depth & measurements \\
\midrule
{A} & 3,211 & 95,844  & 5,758 & 246 & 90\\
\midrule
{B} &2,580 & 62,598 & 3,820 & 176 & 70\\
   & 184 & 16,610 & 1,080 & 100 & 36\\
\midrule
{E} & {1,801} & {78,916} & {5,346} & {202} & {84}\\
& {8,878} & {58,946} & {3,566} & {142} & {68}\\
\bottomrule
\end{tabular}
\end{table}

\subsubsection{Subspace diagonalization}
\label{sec:hpc}

We perform the subspace diagonalization step of ExtSQD and TrimSQD on {Fugaku~\cite{9355239}, Miyabi-G~\cite{10740833}, and ROQUO~\cite{riken2026roquo} as shown in Table~\ref{tab:summary}.}
\begin{itemize}
\item Fugaku is a large-scale Arm-based system comprising 158,976 nodes, each equipped with a Fujitsu A64FX processor featuring 48 compute cores and 32 GB of high-bandwidth HBM2 memory with a peak bandwidth of 1,024 GB/s. The Tofu interconnect D (28 Gbps $\times$ 2 lane $\times$ 10 port) interconnects nodes, allowing efficient large-scale distributed-memory parallel computations. 
\item  
Miyabi‑G is a GPU‑accelerated large‑scale supercomputer comprising 1,120 nodes, each equipped with a single NVIDIA GH200 Grace‑Hopper Superchip. Each node integrates a 72‑core NVIDIA Grace (Arm‑based) CPU and an NVIDIA Hopper H100 GPU, connected through a cache‑coherent NVLink‑C2C interface with an effective bandwidth of approximately 450 GB/s. The Grace CPU provides 120 GB of LPDDR5X memory with a peak bandwidth of 512 GB/s, while the H100 GPU is equipped with 96 GB of HBM3 memory delivering up to 4.02 TB/s. All nodes are interconnected via an InfiniBand NDR200 network in a full‑bisection fat‑tree topology.
\item {ROQUO is a GPU-accelerated supercomputer comprising 135 nodes. Each node is an NVIDIA GB200 NVL4 containing four Blackwell GPUs and 2 Grace CPUs connected via NVLink-C2C interconnect. The system has a total of 540 GPUs and 270 CPUs. Nodes are interconnected by NVIDIA Quantum-X800 InfiniBand networking (up to 3.2 Tbps).}
\end{itemize}

\subsection{Measurement methodology}

{\em{Timing (classical HPC)}} We perform timing measurements using \texttt{std::chrono::steady\_clock}. For sections involving MPI communication, we insert \texttt{MPI\_Barrier} at synchronization points and measure the elapsed time on the master rank, ensuring that the measurement covers the entire duration until all processes complete the corresponding operation.
We measure wall-clock TTS (time-to-solution) as the elapsed time from the start of processing on the first fragment to the completion of processing on the last fragment for the corresponding step. Thus, it reflects both the degree of HPC resource usage and the waiting time due to job scheduling. 

{\em{Timing (QPUs)}} The QPU runtime is the value reported by the quantum processor and is defined as the sum of the circuit execution durations and of the time intervals between consecutive circuit executions. It does not include the time required to compile quantum circuits or transmit them over the network. The number of fragments of different sizes is summarized in Table~\ref{tab:QPUs}, along with their QPU runtimes.

{\em{QPU performance}} We employ as performance metrics the coherence times of qubits (i.e. the timescale over which the computing units of a QPU lose information to the environment), and the error rates affecting 1- and 2-qubit gates (i.e. unary and binary qubit operations) and qubit measurement.

{{\em Timing (HQC)} We measure the end-to-end wall-clock time of the automated workflow (Sec.~\ref{sec:workflow-automation}) as the elapsed time from submission of the first fragment generation job to completion of the energy reconstruction job that finishes last.}

\subsection{Setting of experiments}\label{sec:settings-of-exp}

{\em{Software stack}} We perform Hartree-Fock calculations using MPI-parallelized ORCA~\cite{ORCA6}; {EWF fragment construction, subspace diagonalization, and energy reconstruction using our in-house code;} 
classical CCSD, selected CI, and DMRG calculations with PySCF~\cite{sun2020recent}, DICE~\cite{smith2017cheap} (MPI-parallelized) and Block2 (OpenMP-parallelized)~\cite{zhai2023block2}, respectively.

{\em{Hyperparameters}} {We divide fragments in families according to the number of orbitals: $M_c=13$-18, 19-26, 27-33, 34-$M_c^*$.}

{In baseline calculations (A,B) we choose $M_c^* = \max_c M_c$ (i.e. all fragments are treated with HQC).} We use ExtSQD for fragments with $M_c \leq 26$ and TrimSQD for fragments with $M_c \geq 27$ (a choice based on preliminary experiments that show comparable performance for small fragments but a superior time-to-accuracy for TrimSQD on larger fragments, Fig.~\ref{fig:higher_accuracy}). We execute every fragment (job), regardless of family, using the same configuration of 1,152 {Fugaku nodes}.

{In portability calculations (C,D) we choose $M_c^* = \max_c M_c$ (i.e. all fragments are treated with HQC).} We use TrimSQD for all fragments, regardless of family. We scale the number {of Miyabi-G nodes (GPUs)} per fragment with the family size and assign {8, 16, 64, and 64 nodes to the four families}, respectively.

{Throughout baseline and portability calculations (A-D)}, we fix the subgroup size in step~\ref{key-step} to $8$ for both ExtSQD and TrimSQD, and in each subgroup we perform diagonalization with 10,000 configurations. For TrimSQD we set $k_1=10\%$ and $k_2=50\%$. As standard settings, we choose $\varepsilon=\scinot{1}{-6}$ for ExtSQD and $\varepsilon = \scinot{5}{-6}$ for TrimSQD.

In the refinement calculation (E), we choose $M_c^* = 42$ and assign 96, 576, 1,152, and 2,304 Fugaku nodes to the four fragment families.
In the time-to-solution calculation (F), we choose $M_c^* = 45$. The TrimSQD execution parameters used in (F), including the number of ROQUO nodes, recovery loops, and threshold $\varepsilon$, are summarized in Table~\ref{tab:options}.

\begin{table}[h!]
\centering
\caption{Setup of hyperparameters in the time-to-solution calculation F (T4-Lysozyme, mixed basis, diagonalization on ROQUO).}
\label{tab:options}
\renewcommand{\arraystretch}{0.85}
\setlength{\tabcolsep}{3pt}
\begin{tabular}{ccccccc}
\toprule
\header{{$M_c$}} & \header{13-18} & \header{19-26} & \header{27-33} & \header{34-$M_c^*$} \\ \midrule $N_{\mathrm{node}}$ & 1 & 2 & 8 & 16 \\ $N_{\mathrm{recovery}}$ & 5 & 15 & 15 & 20 \\ $\varepsilon$ & $\scinot{1}{-7}$ & $\scinot{1}{-6}$ & $\scinot{2.5}{-6}$ & $\scinot{2.5}{-6}$ \\
\bottomrule
\hline
\end{tabular}
\end{table}

\subsection{Strong scaling analysis}

To evaluate the strong scaling of TrimSQD and SBD-G matrix-vector multiplication (Sec.~\ref{sec:scaling}) and the trade-off between accuracy and execution time in TrimSQD, ExtSQD, and classical solvers (Sec.~\ref{sec:balancing}), we focus on three representative fragments of trypsin, reported in Table~\ref{tab:atoms-for-sscal}.

\begin{table}[h!]
\centering
\caption{Benchmark fragments used for the strong-scaling analysis.}
\label{tab:atoms-for-sscal}
\setlength{\tabcolsep}{8pt}
\begin{tabular}{lccc}
\toprule
\header{Calculation} & \header{Fragment} & \header{Active space} & \header{FCI dimension} \\
\midrule
{C}   & 178  & \actspace{36}{35} & $\approx \scinot{2.06}{19}$ \\
{C}   & 1098 & \actspace{34}{33} & $\approx \scinot{1.36}{18}$ \\
{C}   & 3211 & \actspace{66}{45} & $\approx \scinot{8.27}{20}$ \\
\bottomrule
\hline
\end{tabular}
\end{table}

\section{Performance results}
\label{sec:performance_results}

\begin{table*}[t!]
\centering
\caption{Total and binding energies, defined as $E_{\mathrm{binding}} = E_{\mathrm{bound}} - E_{\mathrm{unbound}} - E_{\mathrm{ligand}}$, of the benchmark molecular systems.}
\label{tab:total-energy}
\setlength{\tabcolsep}{3pt}
\begin{tabular}{llcccrrr}
\toprule
\textbf{{Calculation}} & \textbf{Method} & $\mathbf{E_{\mathrm{bound}}}$ (Ha) & $\mathbf{E_{\mathrm{unbound}}}$ (Ha) & $\mathbf{E_{\mathrm{ligand}}}$ (Ha) & $\mathbf{E_{\mathrm{binding}}}$ (Ha) & $\mathbf{E_{\mathrm{binding}}}$ (kcal/mol) \\
\midrule
C: trypsin, benzamidine 
 & EWF--CCSD      & $-319,413.0227$ & $-319,038.0806$ & $-374.9875$ & $0.0454$ & $28.46$ \\
{STO-3G} & EWF--TrimSQD       & $-319,415.8966$ & $-319,040.9552$ & $-374.9986$ & $0.0572$ & $35.89$ \\
\midrule
D: T4-Lysozyme, $n$-butyl-benzene
 & EWF--CCSD      & $-289,139.2838$ & $-288,756.3865$ & $-382.9139$ & $0.0166$ & $10.45$ \\
{STO-3G} & EWF--TrimSQD       & $-289,141.6090$ & $-288,758.7087$ & $-382.9152$ & $0.0148$ & $9.30$ \\
\midrule
{C: trypsin, benzamidine}
 & {EWF--CCSD}      & {$-319,413.0227$} & {$-319,037.8220$} & {$-374.9926$} & {$-0.2081$} & {$-130.59$} \\
{STO-3G, electrostatic correction} & {EWF--TrimSQD}       & {$-319,415.8966$} & {$-319,040.6832$} & {$-375.0043$} & {$-0.2092$} & {$-131.32$} \\
\midrule
{E: T4-Lysozyme, $n$-butyl-benzene}
 & {EWF--CCSD}      & {$-289,179.5355$} & {$-288,793.9070$} & {$-385.6009$} & {$-0.0275$} & {$-17.31$} \\
{mixed (3-21G, STO-3G) basis} & {EWF--TrimSQD}       & {$-289,182.0538$} & {$-288,796.4222$} & {$-385.5980$} & {$-0.0336$} & {$-21.07$} \\
\bottomrule
\end{tabular}
\end{table*}

\begin{table*}[t!]
\centering
\caption{QPU performance analysis for calculations A, B, E in Table~\ref{tab:summary}. Full-system values reflect median $\pm$ median absolute deviation across the entire QPU. Fragment performance represents the QPU portion used to execute the largest circuit, see Table~\ref{tab:atoms-for-qpu}.}
\label{tab:QPU_props-median}
\setlength{\tabcolsep}{1.5pt}
\begin{tabular}{cccccccccccc}
\toprule
\multirow{2}{*}{\header{Calc.}}
& \multirow{2}{*}{\header{QPU}}
& \multirow{2}{*}{\header{Metric type}}
& \multirow{2}{*}{\header{\makecell{Number\\of qubits}}}
& \multicolumn{2}{c}{\header{Qubit coherence}}
& \multicolumn{2}{c}{\header{One-qubit gate}}
& \multicolumn{2}{c}{\header{Two-qubit gate}}
& \multicolumn{2}{c}{\header{Readout}} \\

\cmidrule(lr){5-6}
\cmidrule(lr){7-8}
\cmidrule(lr){9-10}
\cmidrule(lr){11-12}

& &  & 
& $T_1$ (\si{\micro\second})  & $T_2$ (\si{\micro\second}) 
& Time (\si{\nano\second}) & Error 
& Time (\si{\nano\second})& Error 
& Time (\si{\nano\second})& Error \\
\midrule
{A,C} & \device{c}   & full QPU & $156$ 
& $286\pm56$ & $168\pm89$
& $32$ & $0.022\pm0.006$\%
& $68$ & $0.181\pm0.045$\%
& $2584$ & $0.671\pm0.195$\%\\

& & fragment 3211 & $94$
& $293\pm53$ & $142\pm76$
& $32$ & $0.022\pm0.006$\%
& $68$ & $ 0.176\pm0.038 $\%
& $2584$ & $0.610\pm0.171$\%\\

\midrule
{B,D} & \device{c}   & full QPU & $156$ 
& $280\pm52$ & $158\pm88$
& $32$ & $0.021\pm0.006$\%
& $68$ & $0.190\pm0.054$\%
& $2584$ & $0.555\pm0.165$\%\\

& & fragment 2580 & $74$
& $281\pm54$ & $181\pm101$
& $32$ & $0.020\pm0.006$\%
& $68$ & $0.219\pm0.074$\%
& $2584$ & $0.500\pm0.098$\%\\

\cmidrule{2-12}
& \device{k}   & full QPU & $156$ 
& $230\pm56$ &$96\pm59$
& $32$ & $0.024\pm0.008$\%
& $68$ & $0.217\pm0.070$\%
& $2160$ & $0.745\pm0.336$\%\\

& & fragment 184 & $40$
& $232\pm63$ &$118\pm39$ 
& $32$ & $0.024\pm0.007$\%
& $68$ & $0.192\pm0.037$\%
& $2160$ & $0.653\pm0.287$\%\\

\cmidrule{1-12}
{E} & \device{c} & full QPU & $156$ 
& $246\pm52$ &$161\pm87$
& $32$ & $0.026\pm0.007$\%
& $68$ & $0.163\pm0.042$\%
& $2180$ & $0.476\pm0.134$\%\\

 & & fragment 1801 & $88$
& $251\pm56$ &$181\pm84$ 
& $32$ & $0.026\pm0.007$\%
& $68$ & $0.140\pm0.030$\%
& $2180$ & $0.439\pm0.140$\%\\

\cmidrule{2-12}
& \device{k}   & full QPU & $156$ 
& $184\pm52$ &$97\pm54$
& $32$ & $0.029\pm0.010$\%
& $68$ & $0.215\pm0.087$\%
& $2180$ & $0.976\pm0.448$\%\\

& & fragment 8878 & $72$
& $186\pm48$ &$124\pm77$ 
& $32$ & $0.029\pm0.011$\%
& $68$ & $0.220\pm0.086$\%
& $2180$ & $0.989\pm0.439$\%\\
\bottomrule
\end{tabular}
\end{table*}

\subsection{Results, binding energy}
\label{sec:results}

Table~\ref{tab:total-energy} presents the ground-state energies for the two benchmark protein-ligand complexes, calculated with lower-scaling EWF using a purely classical and a HQC fragment solver (EWF-CCSD and EWF-TrimSQD). We also report the binding energies, computed with the formula in Table~\ref{tab:total-energy}. {For minimal-basis calculations, EWF-CCSD and EWF-TrimSQD yield binding energies that are comparable but too positive although fragment solutions are accurate. For T4-Lysozyme, we employ: (i) a mixed basis set, using 3-21G for the ligand and nearby residues (i.e. with at least one atom within $3 \, \mathrm{\AA}$ from the ligand) and STO-3G for the rest of the system; and (ii) a more robust bath threshold, $\eta = \scinot{1}{-8}$, to selectively increase computational effort and accuracy in the binding region. These refinements yield a negative binding energy.}

{For trypsin, the bound (unbound) system carries a net charge $Q = 0$ ($Q = -1$). The species removed on going from bound to unbound is the benzamidinium ligand, carrying net charge of $Q = +1$ and shifting the electrostatic potential experienced by fragments by a term inversely proportional to their distance from the charged ligand. We thus subtract from the one-body contribution to the correlation energy from a fragment in the unbound system a term of the form $Q \cdot \Delta q_c/r$, where $\Delta \, q_c$ is the change in the fragment's electron number on going from the mean-field to a correlated description and $r$ is the distance from the fragment's central atom to the centroid of the ligand's amidinium group.
We apply the same construction to the isolated ligand with $Q=-1$ and $r$ measured from the centroid of the Asp189 carboxylate residue. This ``electrostatic correction'' yields a negative binding energy. A more systematic investigation of the treatment of long-range charge fluctuations will be addressed in future work.
}

\subsection{QPU capacity} Quantum sampling on \device{c} and \device{k} used up to 94 and 72 qubits (60\% and 46\% capacity, defined as the ratio between in-use and available qubits), i.e. significant portions of the QPU. Performance metrics are summarized in Table~\ref{tab:QPU_props-median}. 
{In calculation A, this performance is sustained for 56 h using \device{c}. In calculations B and E (minimal, mixed basis) for 44 h  and 47 h using \device{c,k}} as detailed in Table~\ref{tab:EWF_cluster_cal_time}.
For each circuit execution, we selected a qubit layout constrained by hardware and circuit connectivity, using device calibration data to preferentially use low-error qubits and couplers. The selected layouts have median 1- and 2-qubit error rates and coherence times typically higher than device-wide medians, indicating that the reported results reflect typical device performance rather than unusually favorable layouts or transient fluctuations. The largest circuit in Table~\ref{tab:atoms-for-qpu} used $94$ qubits and a total 2-qubit execution time of approximately \SI{17}{\micro\second} (\SI{68}{\nano\second} 2-qubit gate time $\times$ $246$ 2-qubit depth). This duration is shorter than the median coherence times of the qubits $T_1$ (\SI{293}{\micro\second}) and $T_2$ (\SI{142}{\micro\second}) over the selected layout.

\subsection{HPC capacity}

\begin{table}[b!]
\centering
\caption{Computational time for EWF fragment calculations}
\label{tab:EWF_cluster_cal_time}
\setlength{\tabcolsep}{4.5pt}
\begin{tabular}{lccrr}
\toprule
{\textbf{Calc.}} & {\textbf{QPU Time$^\dagger$}} & {\textbf{HPC}} & {\textbf{HPC Time}} & {\textbf{Peak Nodes}} \\
 & (\device{c},\device{k}) & & [node-hours] & (\% of HPC) \\
 \midrule
A & (56,0)         & Fugaku   & $369{,}315$          & $152{,}064 \, (95.7\%)$ \\
B & ($44,6$)       & Fugaku   & $308{,}528$          & $152{,}064\, (95.7\%)$ \\
 \midrule
C & samples from A & Miyabi-G & $4{,}281^{\ddagger}$ & $1{,}104 \, (98.6\%)$ \\
D & samples from B & Miyabi-G & $3{,}455^{\ddagger}$ & $1{,}104 \, (98.6\%)$ \\
 \midrule
E & {({36,47})} & {Fugaku} & {$3{,}772{,}926$} & {$152{,}064 \, (95.7\%)$} \\
 \midrule
F$^\P$ & (25,25) & ROQUO & 7,620 & 126 (93.3\%) \\
\bottomrule
\end{tabular}
\\
\scriptsize{
$^\dagger$ Total QPU time in hours,
$^\ddagger$ GPU-hours (Miyabi-G has one GPU per node), \\
$^\P$ ground-state energy for bound
}
\end{table}

\begin{figure*}[t!]
  \centering
  \includegraphics[width=0.95\linewidth]{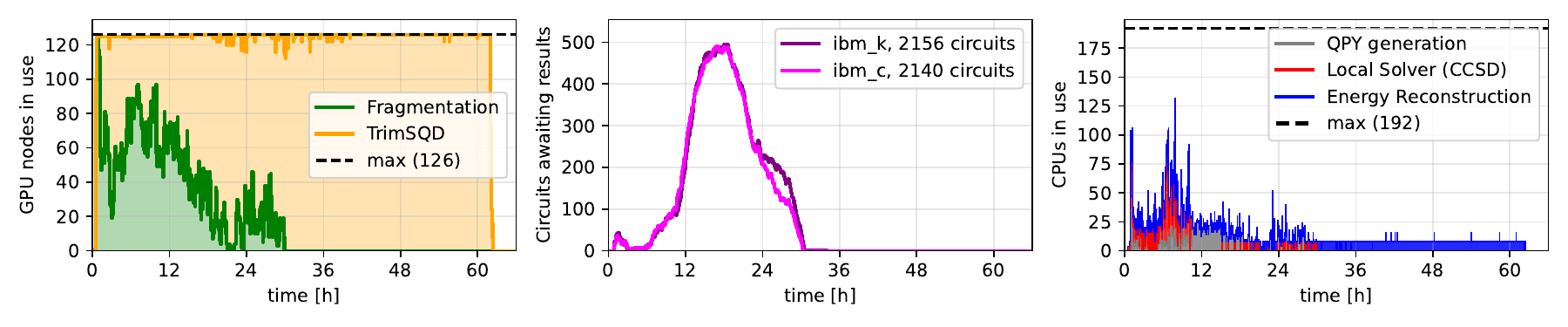}
  \caption{{Timeline of GPU, QPU, CPU (left, middle, right) resource usage for calculation F in Table~\ref{tab:summary} (T4-Lysozyme, $n$-butyl-benzene complex; mixed basis).}}
  \label{fig:tts}
\end{figure*}

\begin{figure*}[t!]
  \centering
  \includegraphics[width=0.95\linewidth]{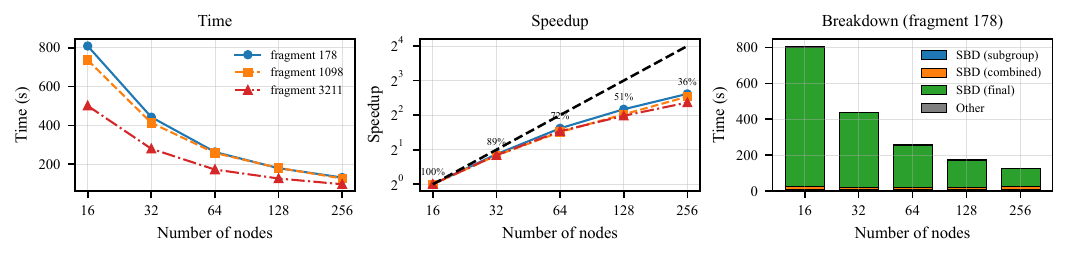}
  \vspace{-6pt}
  \caption{Strong scaling of TrimSQD with baseline threshold $\varepsilon = 5 \cdot 10^{-6}$. In each point, the median of parallel efficiency is annotated (left and center). Elapsed time breakdown in the atom 178 (right).}
  \label{fig:trim-sscal-all}
\end{figure*}

As summarized in Table~\ref{tab:EWF_cluster_cal_time}, {baseline (A,B) calculations on Fugaku sustained near-peak capacity of 152,064 (95.7\%) nodes, while the portability calculations (C,D) on Miyabi-G used up to 1,104 (98.6\%) nodes.
The observed wall-clock elapsed times for the EWF fragment calculations are 78.36 h (A, running as regular non-reserved job, including substantial queueing time) and 4.58 h (B, full system) on Fugaku, and 7.87 h (C, full node) and 9.43 h (D, full node) on Miyabi-G with other experiments running concurrently. The corresponding work-bound runtimes (i.e., excluding queueing and other non-compute overheads) are 2.32 h (A), 1.94 h (B), 3.82 h (C), and 3.08 h (D), respectively.}
{The corresponding total resource consumptions summarized in Table~\ref{tab:EWF_cluster_cal_time} show that, for T4-Lysozyme (system II), going from STO-3G to mixed basis (B,D to E) set corrects the sign of the binding energy with a $\approx 10 \times$ increase in computational cost (i.e. total node-hours).}

The use of two QPUs (\device{c} and \device{k}) and two {supercomputers} (CPU-based Fugaku and GPU-based Miyabi-G) at near-full capacity demonstrates the portability and flexibility of our approach.

{
\subsection{Time to solution}
\label{sec:TTS}

While calculations A-E in Table~\ref{tab:EWF_cluster_cal_time} used a manually coordinated workflow spanning multiple computational resources, Fig.~\ref{fig:tts} reports an additional calculation (Table~\ref{tab:summary}, F) enabled by an automated framework that orchestrates execution across \device{c}, \device{k}, and ROQUO.
Using this automated workflow (\ref{sec:workflow-automation}), we report an end-to-end wall-clock time for a mixed-basis EWF-TrimSQD calculation of the T4-Lysozyme, $n$-butyl-benzene complex (bound), from fragment generation to energy reconstruction, of 62.4 hours.
(i) GPUs performed fragment generation and TrimSQD, with a reservation capacity of 126 nodes. These steps of the computation employed 7,620 node-hours and on average 122.1 nodes (96.9\%, peak 126).
(ii) QPUs executed 4,296 circuits (2,140 on \device{c} and 2,156 on \device{k}) gathering $\scinot{6.08}{8}$ samples over 50.0 h of QPU time. QPU execution did not introduce idle time in the workflow, as GPU computations from other stages proceeded concurrently and GPUs remained active throughout the entire QPU runtime.
(iii) CPUs performed lighter computational tasks, with a reservation capacity of 192 cores. They employed 167 core-hours and on average 2.67 cores (1.39\%, peak 132).
}

\subsection{Scaling}
\label{sec:scaling}

We achieved near-peak capacity on all platforms by parallelizing the subspace diagonalization step of EWF fragment calculations, and we chose the parallelization strategy based on the strong-scaling analysis of TrimSQD and its dominant subroutine, matrix-vector multiplication.

\subsubsection{Strong scaling of TrimSQD}

Fig.~\ref{fig:trim-sscal-all} shows the strong scaling behavior of TrimSQD. Increasing the number of nodes consistently reduces the overall time-to-solution. In the practically relevant regime, TrimSQD exhibits a high parallel efficiency. For example, using $N_{\mathrm{node}} = 64$ nodes as in the main production runs, we achieve a parallel efficiency of approximately $72.5\%$. As the number of nodes increases further, the scaling gradually deviates from the ideal linear behavior. Therefore, for the size of the present problems, moderate node counts provide a favorable balance between parallel efficiency and time-to-solution. In particular, the choice of $N_{\mathrm{node}} = 64$ also reflects practical memory constraints, as it avoids out-of-memory issues while maintaining high efficiency and a substantial reduction in computation time.
Fig.~\ref{fig:trim-sscal-all} (right panel) also shows the breakdown of the execution time in Trim-SQD. From this figure, we observe that the final diagonalization stage dominates the overall runtime. We therefore proceed to analyze the scaling behavior of the diagonalization itself.

\subsubsection{Strong scaling of matrix-vector multiplication}
\label{sec:matvec}

\begin{figure}[t!]
  \centering
  \includegraphics[width=0.95\linewidth]{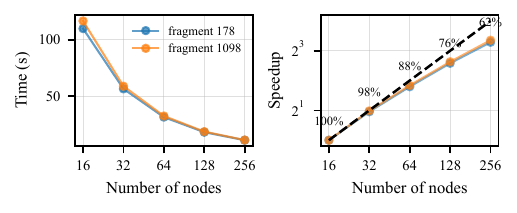}
  \vspace{-6pt}
  \caption{Strong scaling of matrix--vector multiplication ($N_{\mathrm{sub}} = 2^{32})$.}
  \label{fig:sscal-diag}
\end{figure}

Fig.~\ref{fig:sscal-diag} shows the strong scaling behavior of matrix-vector multiplication, the dominant computational kernel in subspace diagonalization.
Here, $N_\mathrm{sub} = 2^{32} \approx \scinot{4.29}{9}$ denotes the dimension of the subspace to be diagonalized. 
The execution time of a single matrix-vector multiplication decreases significantly as the number of nodes increases.
At larger node counts, the scaling gradually becomes sublinear.
However, it is important to note that the execution time per matrix-vector operation reaches approximately $11\,\mathrm{s}$ or less, even for a subspace of dimension $N_{\mathrm{sub}} = \scinot{4.29}{9}$.
Therefore, for the size of the present problems, increasing the count of nodes beyond a certain point yields diminishing returns in terms of time reduction per operation.

The strong scaling of TrimSQD determines the optimal parallelization of subspace diagonalizations across a large-scale HPC platform, ultimately allowing for the scalability of EWF fragment calculations to the full Fugaku and Miyabi-G.

\subsection{Accuracy}
\label{sec:balancing}

Strong scaling is important, but performance gains are meaningful only if the underlying EWF fragments are accurately treated. We thus assess the accuracy of the EWF fragment energies.

\subsubsection{Comparison between TrimSQD and ExtSQD}

First, we compare the accuracy of TrimSQD and ExtSQD, to assess the effectiveness of the innovations reported in Sec.~\ref{sec:innovations}. Fig.~\ref{fig:higher_accuracy} shows the energy differences between TrimSQD and ExtSQD for the 403 non-water fragments with large molecular-orbital counts ($M_c \geq 27$) used in the calculation of $E_{\mathrm{bound}}$ for the trypsin complex.
Since lower energy corresponds to higher accuracy, a negative energy difference indicates that TrimSQD outperforms ExtSQD: indeed, TrimSQD achieves higher accuracy than ExtSQD for almost all fragments despite the use of threshold settings favorable to ExtSQD ($\varepsilon= \scinot{5}{-7}$ for ExtSQD and $\varepsilon = \scinot{2.5}{-6}$ for TrimSQD). Moreover, the advantage of TrimSQD becomes more pronounced as the fragment size (measured by the FCI space dimension) increases.
The TrimSQD/ExtSQD accuracy comparison confirms and quantifies the benefit of the innovations in subspace diagonalization for fragment solution.

\subsubsection{Accuracy-time tradeoff in TrimSQD and ExtSQD}

\begin{figure}[h!]
  \centering
  \includegraphics[width=0.90\linewidth]{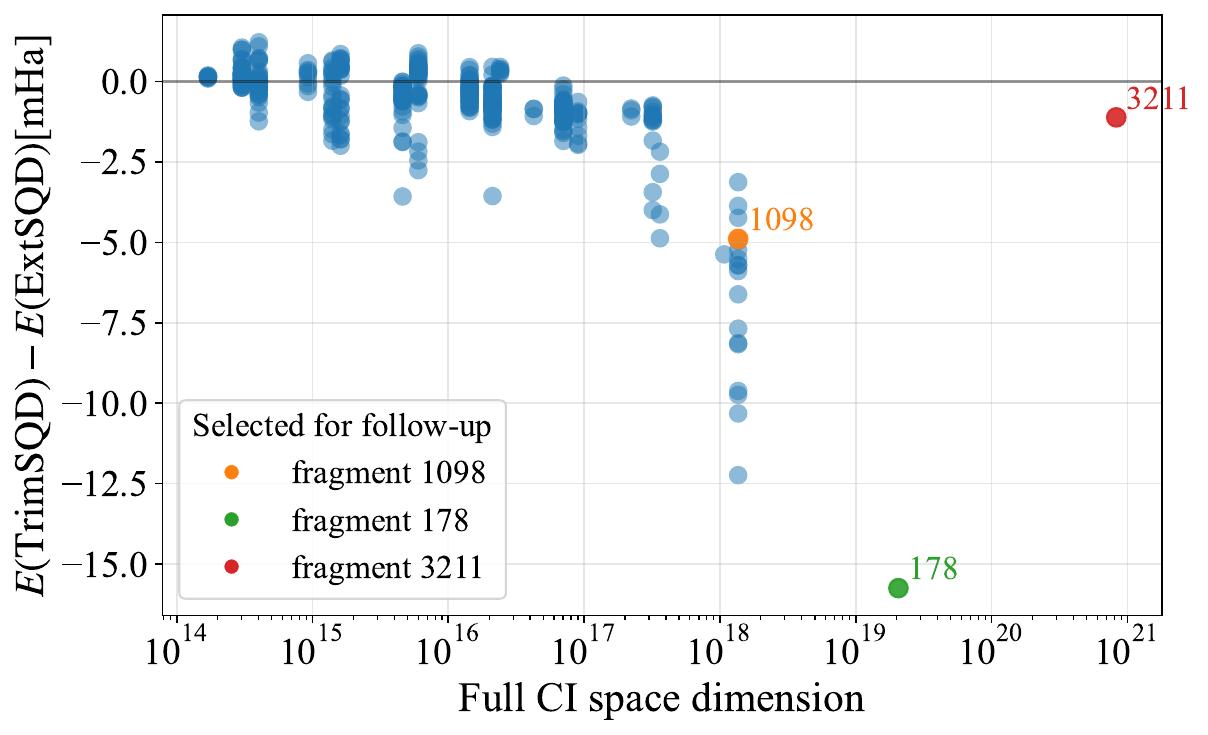}
  \caption{Accuracy improvement of TrimSQD over ExtSQD.
  }
  \label{fig:higher_accuracy}
\end{figure}

\begin{figure*}[t!]
  \centering
  \includegraphics[width=0.96\linewidth]{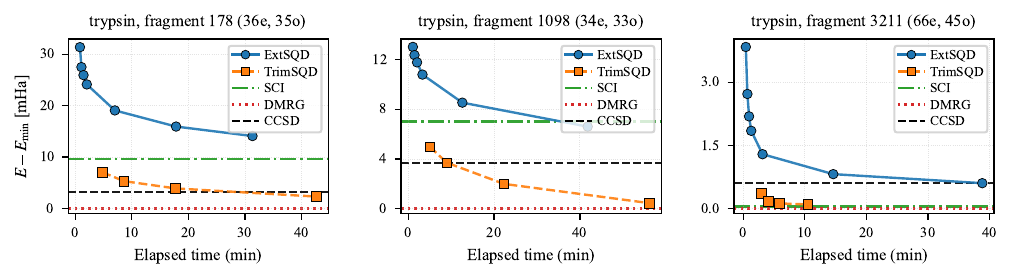}
  \vspace{-12pt}
  \caption{Accuracy-time tradeoffs. ExtSQD and TrimSQD used 128 GPU nodes.}
  \label{fig:Try-acc-time-tradeoff}
\end{figure*}

Second, we perform a time-to-accuracy study comparing TrimSQD with ExtSQD and classical methods, for the three representative fragments highlighted in Fig.~\ref{fig:higher_accuracy} and Table~\ref{tab:atoms-for-sscal}.
As shown in Fig.~\ref{fig:Try-acc-time-tradeoff} and detailed in Table~\ref{tab:best_energies}, TrimSQD consistently achieves high accuracy with substantially reduced time-to-solution compared to ExtSQD and specific CPU-based open-source implementations of SCI and DMRG without particularly developed parallelization strategies. A major factor behind this strong performance is the implementation of SBD-G: unlike conventional SCI, our code does not use storage of the $H_{\bts{x}\bts{y}}$ matrix and distributed storage of the $\Psi_{\bts{x}}$ vectors. As a result, conventional SCI uses significantly more memory, limiting the problem size it can handle. Although the SBD-G design introduces additional arithmetic operations, it remains very efficient; see SCI/TrimSQD data in Table~\ref{tab:best_energies}. Therefore, SBD-G is a key factor that enables TrimSQD to achieve high-accuracy large-scale diagonalization while maintaining practical runtime and memory usage. As a consequence, we do not regard the results reported here as basis of a quantum advantage claim, but as indication that (i) HQC methods are becoming more accurate and efficient and that (ii) progress in computational implementation can further enhance the performance of classical methods. In particular, the integration of SBD-G in the workflow of SCI implementations is a target of our future work, with significant potential to reduce {runtime}.

\begin{table}[h!]
\centering
\caption{Best (i.e. lowest) EWF fragment energies by method}
\label{tab:best_energies}
\setlength{\tabcolsep}{2pt}
\begin{tabular}{rrrrrr}
\toprule
\header{Fragment} & \header{Method} & \header{Energy} & \header{$E-E_{\mathrm{min}}$} & \header{Time} & \header{Dimension} \\
 & & \header{[Ha]} & \header{[mHa]} & $\header{[min]}$ & \\
\midrule
178  & SCI     &  $-204.914157$ &  $9.600$ & 314 & $\scinot{6.27}{7}$  \\
     & ExtSQD  &  $-204.909666$ & $14.091$ &  31 & $\scinot{1.88}{10}$ \\
     & TrimSQD &  $-204.921434$ &  $2.323$ &  43 & $\scinot{1.80}{10}$ \\
     & DMRG    &  $-204.923757$ &  $0.000$ & 308 & N/A                 \\
\midrule
1098 & SCI     &  $-191.178708$ &  $7.019$ & 235 & $\scinot{5.39}{7}$  \\
     & ExtSQD  &  $-191.179119$ &  $6.608$ &  42 & $\scinot{1.04}{10}$ \\
     & TrimSQD &  $-191.185290$ &  $0.437$ &  56 & $\scinot{2.34}{10}$ \\
     & DMRG    &  $-191.185727$ &  $0.000$ & 208 & N/A                 \\
\midrule
3211 & SCI     & $-1108.555377$ &  $0.046$ & 450 & $\scinot{8.87}{7}$  \\
     & ExtSQD  & $-1108.554819$ &  $0.604$ &  39 & $\scinot{2.24}{10}$ \\
     & TrimSQD & $-1108.555325$ &  $0.098$ &  11 & $\scinot{4.99}{9}$  \\
     & DMRG    & $-1108.555423$ &  $0.000$ &  84 & N/A                 \\
\bottomrule
\end{tabular}
\end{table}

\section{Implications}

This study presents a substantial leap of capabilities in modeling molecular systems with quantum embedding and HQC fragment solvers, enabling the simulation of two protein-ligand complexes of unprecedented size. As discussed in Sec.~\ref{sec:innovations}, we overcome the limitations that previously confined these simulations to considerably smaller systems by combining a lower-scaling fragment construction with algorithmic and implementation advances in fragment solutions.

As a result, we cross the 12,000-atom barrier and perform fragment calculations with accuracy higher than previously conducted using HQC methods and matching leading classical WF methods. This is enabled by efficiently leveraging present-day pre-fault-tolerant QPUs and exascale supercomputing resources, achieving high parallel efficiency as demonstrated in Sec.~\ref{sec:performance_results}. {This study executes 21,006 quantum circuits employing up to 94 qubits to collect $\scinot{3.0}{9}$ measurement outcomes}, see Table~\ref{tab:QPUs}, making it the most resource-intensive quantum chemistry HQC calculation to date. {The automated end-to-end workflow elevates these HQC calculations from manually managed, one-off demonstrations to a robust and reproducible production capability, enabling routine operation, automated monitoring, fault recovery, and large-scale computational campaigns, and allowing researchers with access to suitable QPU/HPC resources to execute these workflows without extensive manual intervention.}

This progress substantially extends the boundaries of what is computationally feasible in quantum computing and establishes a framework likely to be widely adopted in quantum computing in the foreseeable future. Indeed, early-fault-tolerant QPUs are projected to substantially reduce the impact of decoherence and imperfect implementation of quantum operations, allowing more accurate calculations, ultimately projected to outperform classical methods~\cite{yoshioka2024hunting,goings2022reliably}. However, they are projected to comprise 200 to 2,000 logical qubits, allowing simulations of 100 to 1,000 molecular orbitals, which supports the view that QPUs will be employed to solve Eq.~\eqref{eq:schrodinger} for fragments in quantum embedding.

Looking ahead, our study establishes a practical baseline for future HQC applications in biochemistry and scientific domains featuring systems of comparable size and chemical intricacy. These include the exploration of covalent-binding drugs, the simulation of catalytic reactions and their transition states, and the characterization of phase diagrams in quantum liquids and solids. Each of these applications will provide an opportunity for quantum computation to participate not only in fundamental research but ultimately in practical innovations in technology and industry.

\section{Acknowledgments}

This work was supported in part through computational resources donated by and services provided by: the Institute for Cyber-Enabled Research at Michigan State University, the high-performance computer center at Cleveland Clinic (Cleveland Clinic Research HPC), the Joint Center for Advanced High Performance Computing (JCAHPC), University of Tsukuba and University of Tokyo (Miyabi-G system, via the Large‑scale HPC Challenge Project), and the RIKEN Center for Computational Science (supercomputer Fugaku). KMM acknowledges financial support from the National Science Foundation (NSF) through CSSI Frameworks Grant OAC-2209717 and from the National Institutes of Health (GM130641). 

Research at RIKEN was supported by the Japan Society for the Promotion of Science (JSPS) KAKENHI (Grants JP26K06972 and JP21H04446); the New Energy and Industrial Technology Development Organization (NEDO), Japan (Project JPNP20017); the Japan Science and Technology Agency (JST) through the COI-NEXT program (Grant JPMJPF2221); the Ministry of Education, Culture, Sports, Science and Technology (MEXT), Japan, through the Program for Promoting Research on the Supercomputer Fugaku (Grant MXP1020230411). Additional support was provided by: the UTokyo Quantum Initiative; the RIKEN TRIP initiative (RIKEN Quantum); the Center of Excellence (COE) Research Grant in Computational Science from Hyogo Prefecture and Kobe City through the Foundation for Computational Science.

We gratefully acknowledge valuable interactions with Hanhee Paik and Yuji Sugita. {We gratefully acknowledge valuable feedback on EWF from George Booth.} {We gratefully acknowledge Akira Naruse and Ikko Hamamura for their valuable support in the development of SBD-G.} {We also thank Shinichi Miura and Yuetsu Kodama for their assistance in securing access to the computational resources used in this work.}



\end{document}